\RequirePackage{fix-cm}
\documentclass[smallextended]{svjour3}       
\smartqed  
\usepackage{graphicx}
\usepackage{amsfonts}
\def\lsim{\mathrel{\rlap{\lower4pt\hbox{\hskip1pt$\sim$}}
    \raise1pt\hbox{$<$}}}         
\def\gsim{\mathrel {\rlap{\lower4pt\hbox{\hskip1pt$\sim$}}
    \raise1pt\hbox{$>$}}}         
    
    \usepackage{float}\usepackage[labelfont={small},font=small,subrefformat=parens,caption=false]{subfig}
\begin{document}

\title{What hath Weinberg wrought? 
}
\subtitle{Reflections on what Weinberg's papers on `Nuclear Forces from Chiral Lagrangians' did and did not accomplish}


\author{Daniel R. Phillips}


\institute{D.~R. Phillips\\
              Department of Physics and Astronomy and Institute of Nuclear and Particle Physics\\
              Ohio University\\
              Athens, OH 45701, USA\\
              Tel.: +1-740-5931698\\
              Fax: +1-740-593-0433\\
              \email{phillid1@ohio.edu}           
}

\date{Received: date / Accepted: date}

\maketitle

\begin{abstract}
I discuss selected legacies of Weinberg's application of chiral Lag\-ran\-gi\-ans to nuclear physics: (1) the use of the chiral expansion to organize the interaction of pions and photons with a nucleus; (2) the much-debated question of why and how the potential derived from a chiral Lagrangian should be inserted in the Schr\"odinger equation; (3)  the emergence of ``pionless EFT" as a tool for diagnosing universal correlations that are present in quantum few-body systems of very different sizes, and, perhaps most important of all, (4) an epistemological shift in what is expected of a nuclear-physics calculation.
\end{abstract}

\section{Introduction}
\label{sec:intro}

In 1994 I was a Ph.D. student at Flinders University in Adelaide, Australia. I was lucky enough to receive a Fellowship to travel to the US so I could present my research at the 14th International Conference on Few-body Problems in Physics. The meeting was in Williamsburg, VA, and included an excursion to Colonial Williamsburg and a tour of the soon-to-be-completed Continuous Electron Beam Accelerator Facility. I remember several plenary talks from that conference, but only one parallel-session talk---other than my own.  A huge crowd gathered for the talk of some bloke who had just finished his Ph.D. at the University of Texas at Austin. His name was Bira van Kolck and he was discussing how to calculate the three-nucleon force from chiral perturbation theory~\cite{vanKolck:1994uv}. 

Looking back, this was the time when three-nucleon forces came into focus. It had recently been shown that (local) NN potentials that reproduced NN data with a $\chi^2/{\rm d.o.f.} \approx 1$ fell short of the triton binding energy by 10\%~\cite{Friar:1993kk}. But it was not immediately clear how to extend the  meson phenomenology employed to construct precise NN potentials to the 3N system---especially because ``purely meson theoretic" NN potentials couldn't achieve $\chi^2/{\rm d.o.f.} \approx 1$ to the NN database. The leading 3NFs of the time---Tucson-Melbourne~\cite{Ellis:1984jh}, Urbana-IX~\cite{Carlson:1983kq}---seemed, at least to a young Ph.D. student, a little {\it ad hoc} in their construction. van Kolck's talk caused excitement precisely because it fixed that. It explained why three-nucleon forces were a small, but important, component of the nuclear force. It showed how to construct them in a manner consistent with the NN force. And it pointed out the inevitable presence in the three-nucleon force of short-distance parameters that need to be fit to data. van Kolck's own contribution to this volume contains more of this story~\cite{vanKolck:2021rqu}.

The rest, as they say, is history. The approach championed by van Kolck was, of course, building on the papers of his Ph.D. advisor, Steven Weinberg, that this volume celebrates, It came to be known as ``Chiral Effective Field Theory" ($\chi$EFT) and has become the dominant paradigm in the construction of nuclear forces.  Its ascendancy is 
partly because of its ability to enforce the construction of consistent NN and 3N forces, and partly because it provides a hierarchy of mechanisms in the nuclear force. These  benefits, together with the development of {\it ab initio} methods to tackle the quantum many-body problem have revolutionized approaches to nuclear structure. Jim Friar's closing talk at FB14~\cite{Friar:1994if} included a shout-out for Brian Pudliner's groundbreaking GFMC calulations of $A \leq 6$ systems~\cite{Pudliner:1995wk}. Today {\it ab initio} methods employ $\chi$EFT potentials to predict the properties of nuclei all the way up to Iron~\cite{Stroberg:2019bch} and in some cases even beyond~\cite{Jiang:2020the,Miyagi:2021pdc}. 

Elsewhere in this volume others tell the history of that impressive progress better than I could~\cite{Drischler:2021kqh}. In this contribution I instead perversely focus on three aspects of Weinberg's seminal papers that either don't involve nuclear forces or don't really have much to do with Chiral Lagrangians. In Section~\ref{sec:probes} I describe how I came to believe that the Weinberg-van Kolck approach would be tremendously useful for electromagnetic and pionic reactions on few-nucleon systems. Such calculations provided early demonstrations that the approach worked for reactions in which it had no free parameters and so was more than ``just fitting"~\cite{Phillips:1999am,Park:1998cu}. In Section~\ref{sec:whyiterate} I return to Weinberg's original papers to amplify a concern he expressed there: why should we iterate the NN potential, i.e., solve a Schr\"odinger equation containing the NN potential derived in $\chi$EFT? I argue that the original sin of the Nuclear Forces from Chiral Lagrangians program was a failure to properly work through this issue: ``compute the potential and just stick it in the Schr\"odinger equation" was, and perhaps remains, the attitude of much of the community. So I review what Weinberg's argument in favor of iteration does and does not establish, and briefly summarize subsequent attempts to justify iteration of the NN potential. This brings me to my third topic, which is a ``toy model" that Weinberg used to explicate some features of what happens when one solves the Schr\"odinger equation with a singular potential. This toy model became ``pionless EFT": a huge contribution to the field of few-body physics. Work on the three-body problem in pionless EFT stimulated tremendous interest in Efimov physics and systems near 
unitarity~\cite{Bedaque:1998kg,Bedaque:1999ve,Braaten:2004rn,Hammer:2010kp}. It is also the basis for the ``Halo EFT" that permits rigorous examination of several systems near the driplines~\cite{Hammer:2017tjm}. Pionless EFT has nothing to do with chiral symmetry, but it remains the only EFT for nuclear physics in which renormalization has been completely and carefully carried out in the two-, three-, and four-nucleon systems. As such it presents a path through some of the issues that have bedeviled the program to systematically obtain Nuclear Forces from Chiral Lagrangians. I conclude with some comments about what makes $\chi$EFT (and pionless EFT) better approaches to few-nucleon systems than the ones that were in use pre-Weinberg.

I want to be clear that this paper contains no new research, and probably nothing I haven't said before somewhere. But I am grateful to the Editors of this Volume for giving me a chance to reflect on the intellectual influence of Weinberg's papers, to reminisce about my own small part in this story, and to articulate what I see as some of the unfinished business of Weinberg's work. 

\section{Electromagnetic and pionic reactions on deuterium}
\label{sec:probes}

My favorite ``$\chi$PT for nuclear physics" paper by Weinberg is not thirty until next year. It's ``Three-body interactions among nucleons and pions", the oft-overlooked third of the trio of nuclear-physics-transformational papers he wrote between 1990 and 1992~\cite{Weinberg:1990,Weinberg:1991,Weinberg:1992}. This paper has ``only" 497 citations on INSPIRE as I write this sentence (compared to the massive 1,370 and 1,302 citations for the two ``Nuclear forces from chiral Lagrangians" papers). In Ref.~\cite{Weinberg:1992} Weinberg does two very interesting things. First, he establishes that the nominally leading (these days we would say ``NLO" or $\mathcal{O}(Q^2)$)~\footnote{Throughout I denote the $\chi$EFT expansion parameter by $Q$, and intend it to be the standard $Q \equiv \frac{(p,m_\pi)}{\Lambda_b}$ where $\Lambda_b$ is the $\chi$EFT breakdown scale, assumed to be of order $m_\rho$.} contribution to the $\chi$PT three-nucleon force vanishes, something that was worked out in more detail by van Kolck in Ref.~\cite{vanKolck:1994yi}. But Weinberg then decides that this makes pion-nucleon-nucleon interactions more interesting. So he calculates the leading $\chi$PT diagrams for those and uses them to obtain a prediction for the pion-deuteron scattering length. 

It was probably partly that pion-deuteron scattering was my Ph.D. project, but this paper made a big impression on me. It helped that $\chi$EFT works remarkably well for the pion-deuteron scattering length. Compared to the complicated formalism I was using to try and analyze the problem~\cite{Phillips:1995da,Phillips:1996ed} the simplicity and efficacy of the $\chi$PT approach was a revelation. Weinberg's prediction was $-0.05$ $m_\pi^{-1}$, which is certainly consistent with the experimental number (at the time) of $-0.056 \pm 0.009$ $m_\pi^{-1}$. 
The success turns out to be a bit of a cheat, because it's driven by the large contribution from the second-order term in the multiple-scattering series, which gives almost the full answer~\cite{EricsonWeise}. But for me the striking thing was that $\chi$PT provides a systematic approach to the pion-deuteron scattering length. Writing
\begin{equation}
a_{\pi d}=\langle \psi_d|t_{\pi N} + t_{\pi NN}|\psi_d \rangle.
\label{eq:3B}
\end{equation}
and computing the $\chi$PT order of the two-body ($t_{\pi N}$) and three-body ($t_{\pi NN}$) pieces of the result, provides an organized expansion for $a_{\pi d}$.  
In particular Eq.~(\ref{eq:3B}) is a controlled way to calculate the ``nuclear" corrections to the first term, whose matrix element evaluates to the isoscalar pion-nucleon scattering length, i.e., $\chi$EFT provides an improvable approach to $t_{\pi NN}$.

When I visited Seattle in 1996 Bira van Kolck and I spent some time trying to work out how to extend the $\pi$d scattering calculation above threshold. In the end, we didn't pursue that avenue. But I ended up in a collaboration that computed next-order corrections to Weinberg's calculation, facilitating tests of $\chi$PT results for the isoscalar pion-nucleon scattering length~\cite{Beane:2002wk}.  It turns out that there are several wrinkles to power counting for this process~\cite{Baru:2010xn,Baru:2011bw,Baru:2012iv}, but what Ref.~\cite{Weinberg:1992} did---at least for me---was make it clear that EFT arguments can organize processes on deuterium---and by extension processes on other light nuclei.
 
Then there is Compton scattering from deuterium. Because it involves two photons the model calculations of it that had been done by the mid-90s were quite complicated~\cite{Weyrauch:1983zft,Levchuk:1999zy}. Weinberg's ``Three-body interactions" paper provided a way out of the thicket . Once again we write:
 \begin{equation}
T_{\pi d}=\langle \psi_d|t_{\gamma N} + t_{\gamma NN}|\psi_d \rangle,
\label{eq:Compton}
\end{equation} 
where $t_{\gamma NN}$ includes mechanisms in which the photon interacts with two nucleons. Moreover, the presence of a momentum conserving delta function on the spectator nucleon, i.e., the fact that $t_{\gamma N}$ is really $t_{\gamma N} \otimes \mathbb{I}$ when embedded in the $\gamma NN$ Hilbert space, means that $t_{\gamma N}$ dominates over $t_{\gamma NN}$. This observation, made in Ref.~\cite{Friar:1996zw}, specifies the size of corrections to an impulse approximation (i.e. single-nucleon interactions only) treatment of reactions with external probes. 

\begin{figure}[h]
 \centering
   \includegraphics[width=0.7\linewidth]{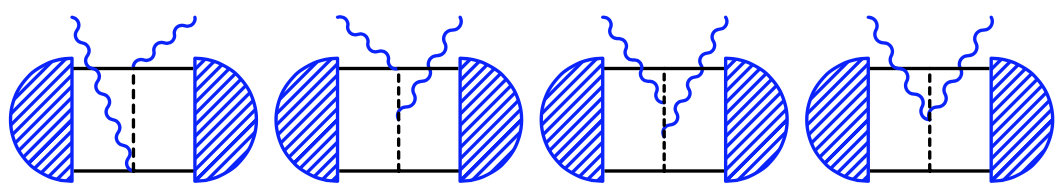}\\
\includegraphics[width=0.7\linewidth]{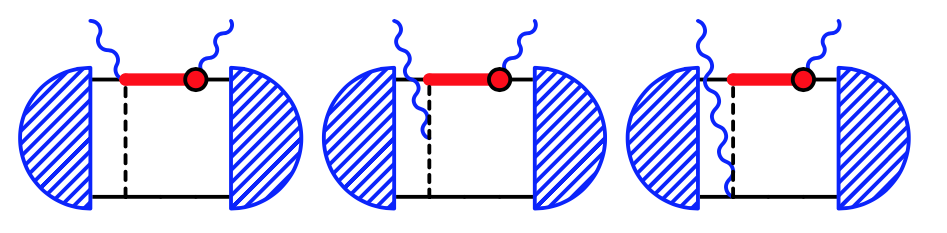} 
\caption{Diagrams that contribute to $\gamma$d scattering at next-to-leading order (top row) and next-to-next-to-leading order (bottom row) in $\chi$EFT. The shaded semi-circles are deuteron wave functions. The red horizontal lines represent nucleon propagators whose power counting changes depending on whether they carry photon energy $\omega$ of order $m_\pi$ or of order $m_\pi^2/M_N$. Figure adapted from Ref.~\cite{Griesshammer:2012we}.}
 \label{fig:Compton}
\end{figure}

 In $\gamma$d scattering those corrections due to $t_{\gamma NN}$ come at next-to-leading order in the $\chi$EFT expansion. There are nine diagrams that enter $t_{\gamma NN}$ at that order (the five in the upper row of Fig.~\ref{fig:Compton} plus diagrams with initial- and final-state photons interchanged). They are the two-nucleon analogs of the diagrams that generate the dominant contributions to the dipole electric and magnetic polarizabilities of the nucleon~\cite{Beane:1999uq}. But organizing the problem according to Eq.~(\ref{eq:Compton}) only makes sense if the diagram is irreducible in the Weinbergian sense. In my Ph.D. I had learnt a topographical notion of irreducibility~\cite{Phillips:1993gg}. I had to unlearn it when I started working on $\gamma$d scattering and realize that for Weinberg the diagrams is irreducible if it has no low-energy NN state in it. Because my collaborators and I only considered a $t_{\gamma NN}$ that was irreducible in this sense the calculations of $\gamma$d scattering we produced in between 1999 and 2005~\cite{Beane:1999uq,Beane:2004ra,Hildebrandt:2004hh} are only valid for photon energies that put the intermediate NN-state far enough off shell that the standard $\chi$PT counting rules apply to the diagrams in $t_{\gamma NN}$, i.e., they are valid for $\omega \sim m_\pi$. As the photon energy goes to zero---or more precisely as it becomes of order $m_\pi^2/M_N$---diagrams such as those in the lower row of Fig.~\ref{fig:Compton} become reducible and one must face the vexed question of how to organize them in the EFT~\cite{Griesshammer:2012we}. Weinberg's answer to that question is the subject of the next section.   
 
 \section{Why iterate?}
  \label{sec:whyiterate}
  
 The leading-order (LO) NN potential in $\chi$PT is an operator whose naive chiral order is $Q^0$:
\begin{eqnarray}
\langle {\bf p}'|V^{(0)}|{\bf p} \rangle&=&C^{{}^3{\rm S}_1} P_{{}^3S_1} + C^{{}^1S_0} P_{{}^1S_0} + V_{1 \pi}({\bf p}'-{\bf p}); \label{eq:V0}\\
 V_{1 \pi}({\bf q})&=&-\tau_1^a \tau_2^a \frac{g_A^2}{4 f_\pi^2}\frac{\sigma_1 \cdot {\bf q} \sigma_2 \cdot {\bf q}}{{\bf q}^2 + m_\pi^2}. \label{eq:VOPE}
\end{eqnarray}
Here $V_{1 \pi}$ is the one-pion-exchange potential, $P_a$ is a projector onto the NN partial wave $a$, and $C^a$ is a low-energy constant (LEC) appearing in the NN chiral Lagrangian. This justifies the dominance of one-pion exchange in nuclear forces; it also predicts the presence of short-distance operators---although only in S-waves. 

Weinberg's papers~\cite{Weinberg:1990,Weinberg:1991} organize $V$ so that the order of corrections to $V^{(0)}$ is determined by naive dimensional analysis (NDA) with respect
to the light scales $(p,m_\pi)$. The leading correction is then  $\mathcal{O}(Q^2)$, and at that order  a number of short-distance operators $\propto {\bf p}^2$ enter, together with several two-pion-exchange diagrams. This counting for $V$ and the progress it has enabled are discussed in other papers in this volume~\cite{Machleidt:2021ggx} as well as in reviews such as Ref.~\cite{Epelbaum:2019kcf}. The use of NDA to organize contributions to the NN potential is not controversial 
for momenta of order $m_\pi$. Once the particle content of the theory is fixed the relative size of the various two-pion, three-pion, etc. mechanisms is dictated by the $\chi$PT power counting.  This gives us control over the long-distance ($r \sim 1/m_\pi$) pieces of the potential.  

But the counting of the short-distance (contact) pieces of $V$ is controversial. It is tied to the unresolved question of whether
the resulting potential should be used as a ``typical NN potential", i.e., inserted into the Schr\"odinger equation and used to solve for the NN wave function. Such a solution is equivalent to iterating the potential to all orders, i.e., summing the Born series generated by $V$. But, at a formal level, there is no immediate reason to think the Born series needs to be summed.
In the single-nucleon and meson sectors $\chi$PT works because pion-nucleon and pion-pion interactions are weak at low energies: that's why they can be treated in a perturbative expansion. And indeed naive analysis of powers of momenta tells us that the nuclear potential begins at $\mathcal{O}(Q^0)$, while the (one-body) kinetic-energy operator, $K$, is $\mathcal{O}(Q^{-1})$, and so suggests the nuclear force is perturbative, since $V < K$. 
But if we lived in a universe where nuclear interactions were perturbative then there would be no nuclei, no nuclear physics, and no nuclear physicists who spend twenty-five years arguing about whether $\chi$EFT potentials should be iterated or not. 
 
\begin{figure}[h]
\begin{center}
\includegraphics[width=0.6\linewidth]{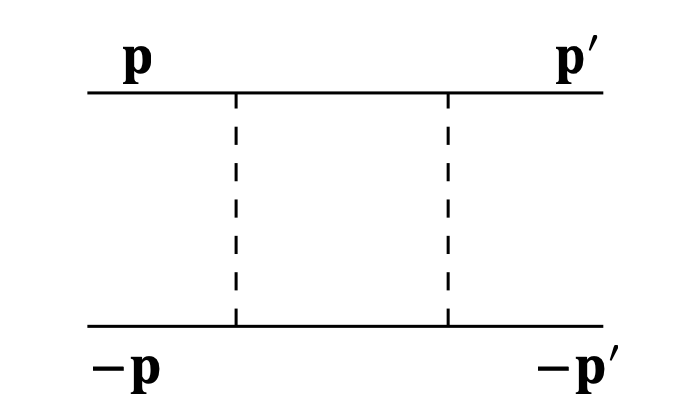} 
\end{center}
\caption{A reducible diagram, involving iteration of the one-pion-exchange potential, see Eq.~(\ref{eq:Titerate}). External lines are labeled with the three-momentum that they carry.}
\label{fig:iterate}
\end{figure}

Weinberg justified the insertion of the potential (\ref{eq:V0}) in the Schr\"odinger equation by noting that diagrams with a low-energy NN state, such as Fig.~\ref{fig:iterate}, produce a singularity if they are treated using the standard heavy-baryon $\chi$PT propagator. If we take the propagator for a single nucleon of energy $p_0$ to be $i/p_0$ then the propagator for an NN state of energy $E$ is $i/E$, and as $E \rightarrow 0$ this blows up. The solution to the problem of this $1/E$ singularity was articulated in the context of non-relativistic QED in the 1980s~\cite{Caswell:1985ui}: if $p_0 \rightarrow 0$ then the standard heavy-baryon approximation that $p_0 \gg \frac{{\bf p}^2}{2 M_N}$ no longer applies, and the single-nucleon propagator becomes $\frac{i}{p_0 - {\bf p}^2/(2 M_N)}$. Doing the integral over the relative energy of the two nucleons yields a two-nucleon propagator that is the standard non-relativistic Schr\"odinger equation Green's function, i.e.,
\begin{equation}
\langle {\bf p}'|G_0(E)|{\bf p} \rangle=\frac{i}{E^+ - \frac{{\bf p}^2}{M_N}} \delta^{(3)}(p-p').
\label{eq:propagator}
\end{equation}
The ``pinch singularity" that Weinberg used to justify iteration is thus ameliorated by the nucleon kinetic energies: reducible diagrams---diagrams with low-energy NN intermediate states---do not actually diverge. 

They are enhanced though: for $E \sim m_\pi^2/M_N$ and $|{\bf p}| \sim m_\pi$ the propagator (\ref{eq:propagator}) is of order $M_N/m_\pi^2$, and so is markedly larger than the typical HB$\chi$PT propagator $i/q_0$ that we take to be $\sim 1/m_\pi$. But the kinematic situation is also different: in a single-nucleon HB$\chi$PT diagram the heavy particle is usually off shell by an amount of order $m_\pi$, but here we are dealing with nearly-on-shell nucleons. 

Inserting the propagators for non-relativistic nucleons and doing the integral over the NN relative energy converts the diagram in Fig.~\ref{fig:iterate} to the expression
\begin{equation}
-i \langle {\bf p}'|T_{\rm iterate}(E)|{\bf p} \rangle=\int \frac{d^3 p''}{(2 \pi)^3} (-i) V_{1 \pi}({\bf p}''-{\bf p}') \frac{i}{E^+ - \frac{{{\bf p}''}^2}{M_N}} (-i) V_{1 \pi}({\bf p}''-{\bf p}),
\label{eq:Titerate}
\end{equation}
where $V_{1 \pi}$ is the one-pion-exchange potential of Eq.~(\ref{eq:VOPE}). 
Assuming that we are dealing with energies of order $m_\pi^2/M_N$ we expect that the integral will be dominated by momenta of order $m_\pi$. The (naive) chiral order of the diagram is then $Q^3 \times Q^0 \times M_N/Q^2 \times Q^0$. Keeping factors of $\pi$ we find  $T_{\rm iterate} \sim \frac{M_N}{4 \pi} Q$, with the dimensions are made up by the factors of $1/f_\pi^2$ in $V_{1 \pi}$. The nominal order of $T_{\rm iterate}$ is thus $Q^1$, while that of $V_{OPE}$ is $Q^0$. So, once again, we are faced with the dilemma that $\chi$PT implies that nuclear reactions should be able to be computed in perturbation theory---a prediction that is in clear contradiction with reality. $V^{(0)}$ is called ``the strong force" for a reason, after all. 

Weinberg's route out of the dilemma, which has also been adopted by Epelbaum and collaborators (see, e.g., Ref.~\cite{Epelbaum:2004fk}), is to count $M_N \sim \Lambda_b/Q$, i.e., say that the scale $M_N$ is larger than the natural breakdown scale of $\chi$PT and so iterates of one-pion exchange---or any other nuclear potential---have additional enhancements which make them as big as their Born-approximation counterparts. I have always found this to be rather unsatisfactory. $M_N$ falls nicely in between $m_\rho$ and $4 \pi f_\pi$, so the scale assignment $M_N \sim \Lambda_b$ employed in the single-baryon sector seems reasonable. Indeed, taking $Q \approx m_\pi/\Lambda_b$ Weinberg's scaling would seem to require an $M_N$ of order a few GeV.  In the single-nucleon sector taking $M_N \sim \Lambda_b$ is not only numerologically satisfying but seems to provide phenomenological success. Changing the assignment to make $M_N$ larger by a factor of $1/Q$ means that all $\chi$PT amplitudes computed in the single-nucleon sector should, in principle, be re-organized before being employed in a few-nucleon calculation. Over the last five years the Bochum group has been carrying out this reorganization, thereby restoring the original $\chi$EFT promise of a consistent treatment of chiral single-nucleon and multi-nucleon amplitudes~\cite{Siemens:2016jwj,Filin:2020tcs}. But this leaves me wondering what the counting of $M_N$ for amplitudes in the single-baryon sector really should be. Reassigning the counting of the nucleon mass to justify iteration says that inside nuclei the nucleon kinetic energy---indeed any nucleon motion---is smaller than anticipated: the reassignment pushes the kinetic-energy operator $K$ to higher order in the $\chi$EFT expansion. What if there's a different---and better---reason to iterate; one that involves acknowledging $V$ is larger than naive chiral counting suggests?
  
In 1998 Kaplan, Savage, and Wise took the argument that iterates of one-pion exchange are smaller than $V_{OPE}$ itself to its logical and ostensibly absurd conclusion~\cite{Kaplan:1998tg,Kaplan:1998we}. They attempted a formulation of nuclear physics in which one-pion exchange was perturbative. In particular, they identified a scale 
\begin{equation}
\Lambda_{NN} \equiv \frac{16 \pi f_\pi^2}{g_A^2 M_N} \approx 300~{\rm MeV}
\end{equation}
below which one-pion exchange can be treated in perturbation theory and above which it must be iterated. This analysis was later confirmed by Birse, using renormalization-group methods~\cite{Birse:2005um}. Qualitatively $\Lambda_{NN}$ is the momentum at which one-pion exchange becomes strong.  It's important to note that $\Lambda_{NN}$ varies only weakly with quark mass, so one can't force it to low or high values by considering the chiral limit when $m_q \rightarrow 0$. Instead, it's numerical value is a result of the interplay of certain hadronic matrix elements. 
In certain hadronic molecules where one-pion exchange is also present the corresponding hadron matrix elements are small enough that $\Lambda_{NN}$ is a high scale and one-pion exchange is perturbative~\cite{Fleming:2007rp,Valderrama:2012jv}. 
But in the NN system $\Lambda_{NN}$ presents a finite scale that's well below $\Lambda_b$ even when one considers the chiral limit. 
Indeed Long and Yang have pointed out that numerical pre-factors can render $\Lambda_{NN}$ lower in some channels, e.g., in the ${}^3$P$_0$ channel of NN scattering it is of order 150 MeV, i.e., $\Lambda_{NN} \sim m_\pi$~\cite{Long:2011qx}. This means that, at least there, one-pion exchange should be iterated already for momenta of order $m_\pi$. And one could argue that $\Lambda_{NN}$ is close enough to $m_\pi$ even in the ${}^3$S$_1$-${}^3$D$_1$ channel that iteration is justified. 

To incorporate the fact that pion-exchange is a strong force we identify $\Lambda_{NN}$ as a low scale in the EFT. $V_{OPE}$ then formally becomes $\mathcal{O}(Q^{-1})$. Now we are relying on a scale hierarchy 
\begin{equation}
m_\pi \sim p \lsim \Lambda_{NN} \ll \Lambda_\chi \lsim M_N.
\label{eq:hierarchy}
\end{equation} 
For summaries of the consequences of this for the organization of the NN potential see Refs.~\cite{Birse:2009my,Phillips:2013fia}. 
Equation~(\ref{eq:hierarchy}) is quite a few scales, and that's before we discuss the need to integrate out ``radiation pions" that are associated with the scale $\sqrt{M_N m_\pi}$~\cite{Fleming:1999ee,Mondejar:2006yu}. Not to mention the fact that two low-lying hadronic resonances, the $\sigma$ and the $\Delta$, also play a significant role in the NN force. 

I think we probably live in an unfortunate~\footnote{Or maybe fortunate? If the NN force was not this complicated perhaps we wouldn't be here?} universe where these scales are really too close together for there to be a straightforward EFT that can be constructed for nuclear forces. There's a reason the nuclear-force problem occupied so many person-hours between 1932~\cite{Bethe:1953}, when Chadwick discovered the neutron, and 1990, when Weinberg wrote Ref.~\cite{Weinberg:1990}. Weinberg's simplification of that problem launched $\chi$EFT and revolutionized the study of nuclear forces, but it did so partly by eliding the many scales that play a role in nuclear forces and making the problem as simple as it would be if we lived in a world where $m_\pi$ and $\Lambda_{NN}$ were much smaller than all other relevant momentum scales. 

\section{Weinberg's toy model begets a cross-sub-field research program}
\label{sec:pionless}

All of us who have sat down and numerically solved a regulated version of the Schr\"odinger equation with the potential given by Eq.~(\ref{eq:V0}) know that the value obtained for the low-energy constants $C_{{}^3S_1}$ and $C_{{}^1S_0}$ depends on the scale of regularization and the renormalization scheme adopted, see Refs.~\cite{Phillips:1999am,Beane:2001bc,PavonValderrama:2004nb,PavonValderrama:2005gu,Yang:2007hb} for explicit examples. The renormalization group describes the evolution of these LECs as the renormalization scale, $\mu$ is changed~\cite{Birse:2005um}. Suppose, then, that we evolve the renormalization scale down to a value $\mu \lsim m_\pi$. At this resolution scale pions are integrated out of the theory; all the physics of s-wave NN interactions resides in the two contact interactions~\footnote{Contact interactions in higher partial waves are also induced~\cite{Nogga:2005hy,Eiras:2001hu}.}. At  resolution scales  $\mu < m_\pi$ the Lagrangian has little to do with chiral symmetry, since the connection to the $m_q \rightarrow 0$ limit is lost for values of $\mu$ this low. With one-pion exchange integrated out what we have left is the leading-order Lagrangian of ``pionless EFT".

That Lagrangian, which includes only the two contact interactions present in Eq.~(\ref{eq:V0}), was written down by Weinberg in his seminal papers. The NN contact interactions are of chiral order 0, which makes an interesting contrast with pion-nucleon interactions. The pseudo-Nambu-Goldstone boson nature of the pion means that pion-nucleon interactions always involve a power of the pion momentum or the quark mass. All pion interactions have a chiral order of at least 1. Nucleon-nucleon interactions are not protected by chiral symmetry in this way: the NN amplitude has no Adler zeros. Chiral symmetry predicts that NN interactions are stronger then $\pi$N or $\pi \pi$ interactions.

\begin{figure}[h]
\begin{center}
\includegraphics[width=0.5\linewidth]{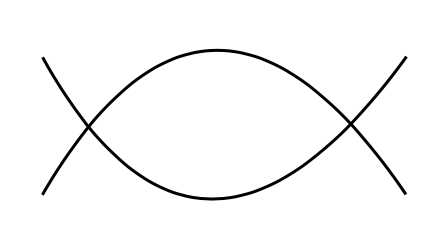} 
\end{center}
\caption{The diagram in which a NN contact interaction is iterated, see expression (\ref{eq:iteratecontact1}).}
\label{fig:bubble}
\end{figure}

But, as already discussed, an interaction of chiral order 0 is not enough to justify iteration. In the case of contact interactions the argument is even more straightforward than the one given in the previous section. Let's consider an RG scale $\mu$ of order $m_\pi$. If $C$ is natural with respect to $\mu$ and $m_\pi$, we have:
\begin{equation}
C \sim \frac{4 \pi}{M_N m_\pi},
\end{equation}
where the factor of $M_N$ in the denominator is a generic feature of LEC scalings in non-relativistic EFTs~\cite{Luke:1996hj}. Meanwhile, computing the diagram in Fig.~\ref{fig:bubble} yields
\begin{equation}
T_{\rm iterate~contact}(E)=C^2 \int d^3 p \frac{1}{E^+ - \frac{p^2}{M}}.
\label{eq:iteratecontact1}
\end{equation}
The integral is linearly divergent. If evaluated using dimensional regularization and minimal subtraction (DR + MS) one obtains a finite result~\cite{Kaplan:1996xu}:
\begin{equation}
T_{\rm iterate~contact}(E)=-C^2 \frac{i M k}{4 \pi} \Rightarrow |T_{\rm iterate~contact}| \sim C \frac{k}{m_\pi}
\label{eq:iteratecontact}
\end{equation} 
where $k=\sqrt{M E}$. This evaluation obscures the presence of the divergence, since the use of DR + MS assumes that all power-law divergences are renormalized by the LECs present at that order and does not keep track of them. What DR + MS makes clear is that the (renormalized) one-loop diagram will be suppressed compared to $C$ itself for momenta $k < m_\pi$, i.e., the momenta for which the EFT is valid. 

Weinberg did two very smart things in Ref.~\cite{Weinberg:1991}. One was to iterate first and worry about what justifies iteration afterwards. He realized that loop graphs involving the contact interaction form a geometric series. If we continue to use DR + MS to evaluate them, then the NN amplitude obtained after summing the series is: 
\begin{equation}
\langle {\bf p}|T(E)|{\bf p}' \rangle=\left(\frac{1}{C} + \frac{i M k}{4 \pi}\right)^{-1}.
\end{equation}
The corresponding $S$-matrix is unitary. By matching to the leading-order effective-range expansion we obtain the renormalization condition for $C$:
\begin{equation}
C_{\rm DR + MS}=\frac{4 \pi a}{M}
\label{eq:CorCR}
\end{equation}
where $a$ is the NN scattering length. This brings me to the second smart thing Weinberg did. Or, more correctly, to something he did {\it not} do. He did {\it not} use DR + MS to evaluate the integral. Instead he only computed the ``renormalized" value of $C$, $C_R$, i.e., the value one gets after absorbing into $1/C$ the divergent part of the integral in Eq.~(\ref{eq:iteratecontact1})~\footnote{I thank Jambul Gegelia for correcting my initial text on this point.}. Because that linear divergence is zero in DR + MS $C_{\rm DR + MS}=C_R$. 

Regardless of whether Eq.~(\ref{eq:CorCR}) is for $C_R$ or $C$ the value on the right-hand side is not natural, since the NN scattering lengths of 5.4 fm in the ${}^3$S$_1$ channel and $-23$ fm in the ${}^1$S$_0$ channel are markedly larger than $1/m_\pi$. Indeed, $C \rightarrow \infty$ as $\frac{1}{a} \rightarrow 0$, i.e., as we approach the unitarity limit where $T \sim 1/p$ and no length scales are left in the problem. The fact that $C$ is directly related to an unnaturally large physical scale when DR + MS is employed leads immediately to difficulties organizing higher-order terms in the EFT expansion~\cite{Kaplan:1996xu,Phillips:1997xu}. Weinberg knew to quit when he was ahead; he never did explain how to extend his scheme for computing the NN amplitude with contact interactions to the case of contact interactions with derivatives. 

The pathway out of this difficulty was worked out in the late '90s, some 5-10 years after Weinberg's initial papers. Two-body systems with large scattering lengths correspond to bound states (or anti-bound states) with shallow binding. In this case the kinetic energy $p^2/M$ and the potential energy can both be sizable, but they cancel almost completely to yield an overall energy of the quantum state that is close to zero. Minimal Subtraction obliterates this fine tuning between kinetic and potential energy since the integral that corresponds to the effects of the (virtual) kinetic energy of particles inside the quantum potential has its real part set to zero. If a cutoff regulator is used to compute the integral in Eq.~(\ref{eq:iteratecontact}) the fine tuning is preserved~\cite{Cohen:1996my}. But power counting at the level of the potential/Lagrangian is then unclear~\cite{Beane:1997pk}.

Kaplan, Savage, and Wise's elegant Power-law Divergence Subtraction (PDS) scheme builds in the cancellation of kinetic and potential energy but emphasizes their scheme-dependence and non-observability, since both become dependent on the PDS scale $\mu$~\cite{Kaplan:1998tg,Kaplan:1998we}. In PDS 
\begin{equation}
T_{\rm iterate~contact}(E)=-C^2 \frac{M}{4 \pi} (\mu + i k),
\end{equation}
and 
\begin{equation}
C=\frac{4 \pi}{M}\left(\frac{1}{a} - \mu\right)^{-1}.
\label{eq:CPDS}
\end{equation}
Power counting at the level of the LECs can now be accomplished, provided $\mu$ is chosen to be of order $k$. In that case $C_2(\mu) p^2$ is smaller than $C(\mu)$ by an amount $\sim k/m_\pi$.  
In fact, Eq.~(\ref{eq:CPDS}) is anticipated in Eq.~(23) of Ref.~\cite{Weinberg:1991}, where Weinberg discusses choosing a renormalization point corresponding to $E=-\mu^2/M_N$, Such use of ``subtractive renormalization'' for NN EFT was then championed by Gegelia in Ref.~\cite{Gegelia:1998gn} and several subsequent works. 

Weinberg did point out that (his version of) Eq.~(\ref{eq:CPDS}) and the large values of $a$ in the NN problem imply that the NN system is close to an infra-red fixed point. PDS ``works" because it organizes the EFT as an expansion around that fixed point~\cite{vanKolck:1998bw,Birse:1998dk}. Denoting $R \sim 1/m_\pi$ as the range of the nuclear force the NN amplitude $T$ is simultaneously expanded in $k R$ and $R/a$. That is particularly clear in PDS, but, in fact, $T$ can be expanded in these two small parameters without the need to choose any particular subtraction scheme~\cite{vanKolck:1998bw,Birse:1998dk}. Provided all counterterms needed for renormalization at a particular order in the expansion are present any subtraction scheme gives the same answer for $T$ up to that order~\cite{vanKolck:1998bw}. All subtraction schemes give the same physical amplitude. But some subtraction schemes give more physical amplitudes than others. 

These efforts of the late 1990s to consistently formulate pionless EFT produced a theory whose renormalization and convergence pattern are very well understood. Computations to high-order in the NN system followed rapidly~\cite{Chen:1999tn,Rupak:1999rk,Butler:2001jj}. A lasting consequence of this effort was the realization that the contact interaction $C$ has to be enhanced over its naive chiral order of 0 in order to justify iteration. If $1/a$ is a light scale and $\mu \sim k$ is also a light scale then the $C$ of Eq.~(\ref{eq:CPDS}) is $\mathcal{O}(Q^{-1})$. Together with the counting of loops as $\mathcal{O}(Q)$ this justifies iteration of the potential. Considering $\Lambda_{NN}$ as a light scale can be similarly regarded as promoting one-pion exchange to $\mathcal{O}(Q^{-1})$~\cite{Birse:2005um}.

Because pionless EFT is not specific to nuclear physics it can, in fact, be used to compute any quantum few-body problem with large scattering lengths, e.g., ${}^4$He dimers, trimers, and tetramers.. Of course, calling it ``pionless EFT" is a bit odd in such a circumstance: perhaps ``Short-Range EFT" (SREFT) is better, since ultimately the EFT is based on the scale hierarchy $R \ll |a|$, i.e., it's built for systems where the two-body scattering length is large compared to the range of the inter-particle potential, $R$. 

When the SREFT Lagrangian was applied to the three-nucleon system it was found that a three-body force was required at leading order in the EFT in order to prevent the three-nucleon system collapsing (``Thomas collapse")~\cite{Bedaque:1998kg,Bedaque:1999ve}. This seems to destroy a selling point of chiral EFT, that three-body forces are suppressed compared to two-body forces, and it took some people, myself included, a while to get used to the idea of leading-order three-body forces. But  from a renormalization-group perspective it is not shocking that the size of three-body forces depends on the resolution scale. In systems where $|a| \gg R$ they become a leading-order effect for $\mu \lsim 1/R$. 

The presence of a three-body force at leading order is related to pionless EFT's realization of famous few-body phenomena such as the Efimov effect~\cite{Efimov:1971zz} and the Phillips line~\cite{Phillips:1968zze}.  The latter is a correlation between the doublet nd scattering length and the triton binding energy. It had been observed in three-body calculations with different NN potentials~\cite{Phillips:1968zze,Thomas:1975qfg}. Bedaque, Hammer, and van Kolck showed that this correlation is a straightforward consequence of the need for a three-body force at leading order in pionless EFT. Later work established that the Tjon line---the correlation between three- and four-body binding energies---can also be understood in this way~\cite{Platter:2004zs}. This shows that these famous few-nucleon correlations have nothing to do with QCD. They don't even stem from the one-pion-exchange piece of the nuclear force. Instead, they are driven solely by the proximity of the NN system to the unitarity limit in which $|a| \rightarrow \infty$. And the systematic character of pionless EFT means that we can predict the order at which the correlation will be broken: $\mathcal{O}(R^2/a^2)$ for the Phillips line~\cite{Bedaque:2002yg,Ji:2012nj} and $\mathcal{O}(R/a)$ for the Tjon line~\cite{Bazak:2018qnu}. 

These insights opened up new connections between nuclear physics and other quantum few-body problems. 
SREFT was very profitably applied to cold atomic gases near a Feshbach resonance, where it facilitated the diagnosis of the Efimov effect in systems of ${}^{133}$Cs atoms. In Ref.~\cite{Kraemer:2006} the location of a loss feature as a function of two-body scattering length was predicted based on a universal correlation derived in SREFT~\cite{Bedaque:2000ft,Braaten:2004rn}. Subsequently Hammer and Platter predicted that, in the unitarity limit, each Efimov trimer is associated with two Efimov tetramers, and they predicted the binding energies of those tetramers based on a SREFT calculation~\cite{Hammer:2006ct}. This prediction was confirmed in subsequent experiments in Innsbruck~\cite{Ferlaino:2009zz}. 

Two-body bound states in which $R \ll |a|$ only exist because of quantum tunneling, since the particles spend much of their time outside the potential, i.e., beyond the classical regime. This is an alternative way to understand the regime in which SREFT is applicable: it's the EFT of states for which tunneling is essential to their existence. Particles spend most of their time outside the potential, so its details don't affect them very much. Halo nuclei, in which one or more neutrons or protons have a significant fraction of their wave function outside the mean-field potential generated by the bulk of the nucleons in the nucleus, are an example of this phenomenon. SREFT has found rich applications in those systems~\cite{Hammer:2017tjm}, where it has revealed several universal correlations between observables. The search is on for more of these universal correlations and for manifestations of the Efimov effect in the nuclear context. Further discussion of Halo EFT can be found in the contribution of Capel to this volume~\cite{Capel:2021ejr}.

\section{Weinberg's nuclear-physics legacies: universality, chiral forces, error bars, and a more virtuous explanation of few-nucleon physics}

I find it amusing that a calculation in the second of Weinberg's papers that seems to be there mainly as an example of how to solve the Schr\"odinger equation with a singular potential generated such research activity. It's a great example of the law of unintended consequences of research. That particular (and unintended?) legacy of Weinberg's work ultimately has had wider applicability across few-body physics than the chiral Lagrangians he wrote down and organized to understand the nuclear force. 

But that work on chiral Lagrangians launched literally a 1000 subsequent investigations. For most nuclear physicists the legacy of the papers this volume celebrates is that they transformed the way nuclear potentials were computed. In tandem with advances in scientific computing power and the development of powerful new many-body techniques nuclear potentials derived in $\chi$EFT revolutionized the study of nuclear structure. As a result we are now in the situation where {\it ab initio} calculations with potentials that are the descendants of the ones Weinberg wrote down are carried out for a wide variety of nuclei and a number of different processes.  Nuclear forces derived from chiral Lagrangians provide a good description of a tremendous amount of three-nucleon data~\cite{Kalantar-Nayestanaki:2011rzs,Epelbaum:2019zqc}, reproduce properties of nuclei containing more than 100 nucleons~\cite{Jiang:2020the,Miyagi:2021pdc}, and---when combined with consistently calculated currents---describe electroweak observables in light nuclei with high precision~\cite{Filin:2020tcs,Phillips:2016mov,Ekstrom:2014iya,King:2020pza,Baroni:2021vrc}

But why did the Weinberg way win? After all, the {\it ab initio} calculations I mentioned in the previous paragraph could be (and sometimes are) carried out with with the non-$\chi$EFT AV18~\cite{Wiringa:1994wb} or  CD-Bonn potentials~\cite{Machleidt:2000ge}---or with suitably evolved and softened versions thereof~\cite{Bogner:2003wn,Bogner:2006pc}. Perhaps the biggest influence of Weinberg's papers on nuclear physics is that they became part of a change in the philosophy of nuclear theorists. Many of us were discontented with calculations whose sole justification was "Well, that's what my model says''. $\chi$EFT gave us a better way to calculate.

In $\chi$EFT you don't guess the mechanisms you think will be important, you organize them according to power counting in $Q$. This makes $\chi$EFT predictive in a way that few-nucleon calculations before 1990 were not. Prediction in those models sometimes worked, if it was done with models that were carefully calibrated to the relevant data, see, e.g., the lovely work on pp fusion of Schiavilla, Stoks, and collaborators~\cite{Schiavilla:1998je}. But sometimes it didn't, with my favorite example being the deuteron quadrupole moment~\cite{Phillips:2003jz,Phillips:2006im}. Fitting on-shell NN data gets you some things very well, but not others. In $\chi$EFT that's not a surprise, because ``everything that is not forbidden is allowed": eventually, at some order in $Q$, all mechanisms not forbidden by symmetries enter the amplitude. This means that two-body currents and higher-body forces---mechanisms not constrained by NN data or current conservation---eventually show up any time you calculate something other than NN scattering. But, if using your NN force model and a one-body operator to calculate something doesn't put you on top of the data you no longer get to just shrug your shoulders and say ``I guess my potential doesn't work for that quantity".  $\chi$EFT tells you {\it when} those other mechanisms show up and it tells you {\it how} they are related to other processes. And then it tells you what the residual uncertainty in your calculation is, because you know if you stopped at order $Q^k$ your amplitude is missing stuff that is $\mathcal{O}(Q^{k+1})$. I still remember Rob Timmermans saying, during one of the many heated exchanges at the Trento nuclear-forces workshop in 1999~\cite{Timmermans:1999}: ``Real theorists have error bars". $\chi$EFT gave nuclear theorists the chance to get themselves some error bars. And we are getting better and better at producing realistic error bars for nuclear-physics calculations~\cite{Furnstahl:2014xsa,Zhang:2015ajn,Premarathna:2019tup,Filin:2019eoe,Phillips:2020dmw,Maris:2020qne,Wesolowski:2021cni}. 

Philosophers of science talk about a set of ``Explanatory Virtues". If one theory has more Explanatory Virtues than Theory B it's a better explanation. Matthew van Cleave~\cite{vanCleave} lists the virtues as:
\begin{enumerate}
\item ``Explanatoriness: Explanations must explain all the observed facts.
\item Depth: Explanations should not raise more questions than they answer.
\item Power: Explanations should apply in a range of similar contexts, not just the current situation in which the explanation is being offered.
\item Falsifiability: Explanations should be falsifiable---it must be possible for there to be evidence that would show that the explanation is incorrect.
\item Modesty: Explanations should not claim any more than is needed to explain the observed facts. Any details in the explanation must relate to explaining one of the observed facts.
\item Simplicity: Explanations that posit fewer entities or processes are preferable to explanations that posit more entities or processes. All other things being equal, the simplest explanation is the best. (Occam's Razor.)
\item Conservativeness: Explanations that force us to give up fewer well-established beliefs are better than explanations that force us to give up more well-established beliefs."
\end{enumerate}
Weinberg's papers gave us a way to understand few-nucleon systems that was Explanatory, Deep, Powerful, Falsifiable, and Simple. No wonder that in the late '90s several young post-docs and students, as well as a number of more established scientists, voted with their feet and chose to work on $\chi$EFT. 

As I've tried to emphasize here, a number of rather foundational questions regarding the computation of nuclear forces from chiral Lagrangians remain open. But the clear legacy of Weinberg's papers is that they produced dramatic shifts in the technical direction and epistemological orientation of theoretical work on few- and many-nucleon systems. Nuclear physics hasn't been the same since---and that's a good thing. 
  
\section*{Acknowledgments}
I thank Dick Furnstahl and Bira van Kolck, who read a draft of this manuscript and suggested different and thankfully not contradictory improvements. I also thank Martin Hoferichter and Jambul Gegelia for their insightful comments on an earlier version of my text. I am very grateful to all my collaborators over the past twenty-five years who have helped me wrestle with what it means to treat nuclear physics using EFT. Some of you may recognize our conversations in this manuscript. Even if you don't, please know that every one of you helped hone the views I've expressed here. This research was supported by the US Department of Energy under contract no. DE-FG02-93ER40756.


\end{document}